\documentclass[12pt]{article}

\usepackage{amssymb}
\usepackage{amsmath}
\usepackage{amscd}
\usepackage{latexsym}
\usepackage{graphicx}

\usepackage{cite}
\usepackage{enumitem}

\newcommand{\be}{\begin{equation}}
\newcommand{\ee}{\end{equation}}
\newcommand{\Dlt}{\Delta}
\newcommand{\dlt}{\delta}
\newcommand{\prt}{\partial}
\newcommand{\br}{{\bf r}}
\newcommand{\bk}{{\bf k}}
\newcommand{\bfe}{{\bf e}}
\newcommand{\ba}{{\bf a}}
\newcommand{\bp}{{\bf p}}
\newcommand{\bu}{{\bf u}}
\newcommand{\bt}{\beta}
\newcommand{\vp}{\varphi}
\newcommand{\ep}{\varepsilon}
\newcommand{\al}{\alpha}
\newcommand{\ra}{\rightarrow}

\newcommand{\gm}{\gamma}
\newcommand{\om}{\omega}
\newcommand{\Om}{\Omega}
\newcommand{\Gm}{\Gamma}
\newcommand{\dgr}{\dagger}
\newcommand{\lbd}{\lambda}

\newcommand{\cH}{{\cal H}}

\newcommand{\rgl}{\rangle}
\newcommand{\lgl}{\langle}

\begin{document}

\begin{center}

{\Large{\bf Destiny of optical lattices with strong intersite interactions} \\ [5mm]

V.I. Yukalov } \\ [3mm]

{\it Bogolubov Laboratory of Theoretical Physics, \\
Joint Institute for Nuclear Research, Dubna 141980, Russia \\ [2mm]
and \\ [2mm]
Instituto de Fisica de S\~ao Carlos, Universidade de S\~ao Paulo, \\
CP 369,  S\~ao Carlos 13560-970, S\~ao Paulo, Brazil  }
\end{center}

\vskip 5cm

\begin{abstract}

Optical lattices are considered loaded by atoms or molecules that can exhibit 
strong interactions between different lattice sites. The strength of these 
interactions can be sufficient for generating collective phonon excitations 
above the ground-state energy level. Varying the interaction strength makes 
it possible to create several equilibrium three-dimensional phases, including 
conducting optical lattices, insulating optical lattices, delocalized quantum 
crystals, and localized quantum crystals. Also, there can exist finite one- 
and two-dimensional lattices of chains and planes.    

\end{abstract}

\vskip 2mm

e-mail: yukalov@theor.jinr.ru

\newpage

\section{Introduction}

Cold atoms or molecules, loaded in optical lattices, have been intensively studied 
both experimentally and theoretically (see, e.g., review articles 
\cite{Morsch_1,Moseley_2,Yukalov_3} and books \cite{Pethick_4,Ueda_5}). Most often, 
one considers atoms or molecules weakly interacting through local potentials. Such 
systems are appropriately described by a single-band Hubbard model containing two 
constants, a tunneling parameter and on-site interaction strength. In the case of 
a delta-function contact interaction, atoms at different lattice sites interact 
very weakly and one assumes that in the majority of situations these weak interactions 
can be neglected. More important are long-range potentials, such as dipolar potential 
\cite{Ueda_5,Griesmaier_6,Baranov_7,Baranov_8,Gadway_9,Kurn_10,Yukalov_11,Yukalov_12} 
making it necessary to take into account intersite interactions. But dipolar 
interactions also are usually treated as sufficiently weak, otherwise they can make 
the system unstable. In that way, the standard situation is when one treats the 
interaction of particles in an optical lattice as sufficiently weak, such that 
the single-band approximation provides an acceptable description not requiring to 
consider excitations above the lowest energy level. However, taking account of such 
collective excitations above the low-band level can essentially alter the lattice 
properties, for instance, rendering an insulating lattice into a conducting state 
\cite{Yukalov_13,Yukalov_14}.

In the present paper, we pose and answer the general question: What happens if 
particle interactions in an optical lattice increase from very weak, when the 
system properties are mainly governed by the lattice configuration, to very strong, 
when the lattice potential becomes practically negligible? We show that if the 
interaction potential increases to such a strength, that the interactions between 
different lattice sites become crucial and these interactions generate collective 
phonon excitations, then there appears a series of three-dimensional equilibrium 
states, such as conducting lattices, insulating lattices, delocalized quantum 
crystals, and localized quantum crystals. In addition to three-dimensional 
macroscopic lattice states, there can exist low-dimensional finite-size crystalline 
chains and planes.  

Throughout the paper, the system of units is used, where the Planck and Boltzmann 
constants are set to one.  

\section{System Hamiltonian}

Let us consider a system of $N$ atoms or molecules in a periodic external field 
formed by laser beams \cite{Letokhov_15}, which creates an optical lattice described 
by the potential
\be
\label{1}
U_L(\br) = \sum_{\al=1}^d U_\al\sin^2(k_0^\al r_\al) \;   ,
\ee
in which $k_0^\al=2\pi/\lbd_\al$. The lattice vector is defined by the laser 
wavelength with the components $\lbd_\al/2$, where $\al=1,2,\ldots,d$, with $d$ 
being the space dimensionality. This potential is evidently periodic with respect 
to the lattice vector,
\be
\label{2}  
 U_L (\br + \br_0 )  = U_L(\br) \qquad 
\left( r_0^\al = \frac{\lbd_\al}{2}\right)\; .
\ee
The single-atom lattice Hamiltonian is
\be
\label{3}
\hat H_L(\br) = -\; \frac{\nabla^2}{2m} + U_L(\br) \;   .
\ee

The energy Hamiltonian for the system of $N$ atoms, or molecules, interacting 
through a potential $\Phi({\bf r}) = \Phi(-{\bf r})$, reads as
\be
\label{4}
 \hat H = \int \hat\psi^\dgr(\br) \hat H_L(\br) \hat\psi(\br) \; d\br \; + \;
\frac{1}{2} \int \hat\psi^\dgr(\br) \hat\psi^\dgr(\br') 
\Phi(\br-\br') \hat\psi(\br')\hat\psi(\br) \; d\br d\br' \; ,
\ee
with $\hat{\psi}$ being field operators. This Hamiltonian can also be represented 
in a lattice form by expanding the field operators over Wannier functions,
\be
\label{5}
 \hat\psi(\br) =\sum_{nj} \hat c_{nj} w_n(\br-\br_j) \;  .
\ee
Here $n$ labels energy bands, while $j = 1,2,\ldots,N_L$ enumerates lattice sites, 
and ${\bf r}_j$ are atomic positions corresponding to an effective lattice formed 
by both, the optical lattice potential and by atomic interactions. 

Substituting expansion (\ref{5}) into Hamiltonian (\ref{4}), one meets the matrix 
elements for the tunneling parameters
\be
\label{6}
 J_{ij}^{mn} \equiv - 
\int w_m^*(\br-\br_i) \hat H_L(\br) w_n(\br-\br_j) \; d\br \;  ,
\ee
the interaction matrix
\be
\label{7}
 U_{j_1j_2j_3j_4}^{n_1n_2n_3n_4} \equiv \int w_{n_1}^*(\br-\br_{j_1}) 
w_{n_2}^*(\br'-\br_{j_2}) \Phi(\br-\br') 
w_{n_3}(\br'-\br_{j_3}) w_{n_4}(\br-\br_{j_4}) \; d\br d\br' \; ,
\ee
and also the matrix elements
$$
\bp_{jmn}^2 \equiv \int w_m^*(\br-\br_j) (-\nabla^2) w_n(\br-\br_j) \; d\br \; , 
$$
\be
\label{8}
U_L^{mn} \equiv \int w_m^*(\br-\br_j) U_L(\br) w_n(\br-\br_j) \; d\br \; .
\ee
Then Hamiltonian (\ref{4}) becomes
$$
\hat H = - \sum_{i\neq j} \; 
\sum_{mn} J_{ij}^{mn} \hat c_{mi}^\dgr \hat c_{nj} \; + \;
\sum_j \; \sum_{mn} \left [ \frac{\bp^2_{jmn}}{2m} + U_L^{mn} 
\right]  \hat c_{mj}^\dgr \hat c_{nj} \; +
$$
\be
\label{9}
+ \; \frac{1}{2} \sum_{ \{j\} } \;  \sum_{ \{n\} } 
U_{j_1j_2j_3j_4}^{n_1n_2n_3n_4} \; \hat c^\dgr_{n_1j_1} \hat c^\dgr_{n_2j_2} 
\hat c_{n_3j_3} \hat c_{n_4j_4}  \; .
\ee

It looks clear that if the interatomic interactions increase to such an extent 
that the optical-lattice potential becomes negligible, the atoms will self-organize 
forming a quantum crystal. Therefore in order to be able to describe these and other 
intermediate states, it is necessary to be based on a general picture unifying these 
limiting situations. For this purpose, it is possible to resort to the description 
employed for quantum crystals by combining a self-consistent approach characterizing 
a low-energy level above which there appear collective phonon excitations 
\cite{Guyer_16,Yukalov_17}. It is also always possible to choose well-localized 
Wannier functions \cite{Marzari_18}, such that the nondiagonal term $U_{ijij}$ be 
much smaller than the diagonal $U_{ijji}$.

For a single band, Hamiltonian (\ref{9}) reduces to the form
$$
\hat H = - \sum_{i\neq j} J_{ij} \hat c_i^\dgr \hat c_j \; + \; 
\sum_j \left( \frac{\bp_j^2}{2m} + U_L \right) \hat c_j^\dgr \hat c_j \; +
$$
\be
\label{10}
 + \; \frac{1}{2} 
\sum_j U_{jj} \hat c_j^\dgr \hat c_j^\dgr \hat c_j \hat c_j \; +
\;   \frac{1}{2} 
\sum_{i\neq j} U_{ij} \hat c_i^\dgr \hat c_j^\dgr \hat c_j \hat c_i \; ,
\ee
in which
$$
J_{ij} = - \int w^*(\br-\br_i) \hat H_L(\br) w(\br-\br_j)\; d\br \qquad 
(i \neq j) \; ,
$$
$$
U_{ij} =  \int | w(\br-\br_i) |^2 \; \Phi(\br-\br') \;
| w(\br' - \br_j) |^2 \; d\br d\br' \; ,
$$
\be
\label{11}
\bp_j^2 = \int w^*(\br-\br_j) (-\nabla^2) w(\br-\br_j) \; d\br \; , 
\qquad  U_L = \int | w(\br) |^2 \; U_L(\br) \; d\br \; .
\ee
This Hamiltonian corresponds to the Hubbard model, if we omit the terms with 
${\bf p}_j^2$, $U_L$, and with the intersite interactions $U_{ij}$. But these 
terms are necessary for considering collective phonon excitations.

\section{Low-energy states}

Low-energy states of a many-body system in a periodic potential can be described 
by using Green function approach. Here we follow the method suggested in Refs. 
\cite{Yukalov_19,Yukalov_20,Yukalov_21}.

We consider the grand Hamiltonian
\be
\label{12}
H = \hat H - \mu \int \hat\psi^\dgr(\br) \hat\psi(\br) \; d\br \;   ,
\ee
with a chemical potential $\mu$. Green functions are defined as the propagators 
\cite{Kadanoff_22}
\be
\label{13}
 G(\br,t,\br',t') \equiv - i 
\lgl \; \hat T \hat\psi(\br,t) \hat\psi^\dgr(\br',t) \; \rgl  ,
\ee
where $\hat{T}$ is a time-ordering operator and the angle brackets imply statistical 
averaging. Green functions satisfy the Dyson equation that in the matrix form can be 
written as
\be
\label{14}
 G = G_0 + G_0 \left( \Sigma - \Sigma_0 \right) G \;  ,
\ee
with $\Sigma$ being the self-energy and $G_0$ and $\Sigma_0$ being a Green function 
and self-energy of a trial approximation. 

For an equilibrium system, the Green functions depend on time difference,
\be
\label{15}
 G(\br,t,\br',t')  = G(\br,\br',t-t') \; ,
\ee
allowing us to define the Fourier transform
\be
\label{16}
 G(\br,\br',\om) = \int_{-\infty}^\infty  G(\br,\br',t) e^{i\om t} \; dt \; .
\ee
Then the Dyson equation (\ref{14}) writes as
\be
\label{17}
  G(\br,\br',\om) =  G_0(\br,\br',\om) + 
\int K(\br,\br'',\om) G(\br'',\br',\om)\; d\br'' \;   ,
\ee
with the kernel
\be
\label{18}
  K(\br,\br',\om) = \int G_0(\br,\br'',\om)\left[ \; \Sigma(\br'',\br',\om) - 
\Sigma_0(\br'',\br',\om) \; \right] \; d\br '' \; .
\ee
Equation (\ref{17}) can be solved by means of an iterative procedure starting with 
the initial approximation $G_0$. Thus the first-order iteration yields 
\be
\label{19}
  G_1(\br,\br',\om) =  G_0(\br,\br',\om)  + 
\int K_1(\br,\br'',\om) G_0(\br'',\br',\om)\; d\br'' \;  ,
\ee
with the kernel
\be
\label{20}
K_1(\br,\br',\om) = \int G_0(\br,\br'',\om)\left[\; \Sigma_1(\br'',\br',\om) - 
\Sigma_0(\br'',\br',\om)\; \right] \; d\br '' \;   .
\ee
    
For the self-energy of zero approximation, it is possible to take the diagonal form
\be
\label{21}
 \Sigma_0(\br,\br',\om) = U_0(\br) \dlt(\br-\br') \; ,
\ee
with a trial potential that is periodic over the lattice vectors,
\be
\label{22}
  U_0(\br + \br_j) = U_0(\br) \; .
\ee
Then the first-order self-energy reads as
\be
\label{23}
   \Sigma_1(\br,\br',\om) =  \dlt(\br-\br') \left[ U_L(\br) + 
\int \Phi(\br-\br'') \rho(\br'')\; d\br'' \right]  \; ,
\ee
where the atomic density is
\be
\label{24}
 \rho(\br) = \lgl \; \hat\psi^\dgr(\br) \hat\psi(\br) \; \rgl = 
\pm i \lim_{\tau\ra +0} \int_{-\infty}^\infty e^{i\om\tau} G(\br,\br',\om) \; 
\frac{d\om}{2\pi} \;  .
\ee
Here the upper sign corresponds to bosons, while the lower, to fermions. The 
kernel (\ref{20}) becomes
\be
\label{25}
 K_1(\br,\br',\om) = G_0(\br,\br',\om) \left[\; U_1(\br) - U_0(\br)\; \right] \; ,
\ee
with the effective potential
\be 
\label{26}
 U_1(\br) = U_L(\br) + \int \Phi(\br-\br') \rho_0(\br') \; d\br ' \;  .
\ee
The Green function, corresponding to self-energy (\ref{21}) can be represented as 
the expansion over Wannier functions
\be
\label{27}
 G_0(\br,\br',\om) = \sum_{nj} G_n(\om) w_n(\br-\br_j) w_n^*(\br'-\br_j) \; ,
\ee
where
\be
\label{28}
G_n(\om) = \frac{1\pm n(\om_n)}{\om-\om_n+i0} \; \mp \; 
\frac{n(\om_n)}{\om-\om_n-i0}
\ee
and 
$$
 n(\om) = \left( e^{\bt\om} \mp 1 \right)^{-1} \qquad 
\left( \bt \equiv \frac{1}{T}\right) \;  .
$$
The frequency $\omega_n = E_n - \mu$ characterizes the energy $E_n$ of the $n$-th 
quantum state, shifted by the chemical potential. The related zero-order approximation 
for the atomic density is
\be
\label{29}
 \rho_0(\br) = \sum_{nj} n(\om_n)\; |\; w_n(\br-\br_j)\; |^2 \;  .
\ee
In that way, the first-order Green function acquires the form 
\be
\label{30}
 G_1(\br,\br',\om) = G_0(\br,\br',\om) + \sum_{mn} \sum_{ij} G_m(\om) G_n(\om)
\Dlt^{mn}_{ij} w_m(\br-\br_i) w_n^*(\br'-\br_j) \;  ,
\ee
in which
\be
\label{31}
 \Dlt^{mn}_{ij} = \int w_m^*(\br-\br_i) [ \; U_1(\br) - U_0(\br)\; ]
 w_n(\br-\br_j) \; d\br \;  .
\ee

The choice of the trial potential (\ref{22}) is not arbitrary but has to be 
defined in such a way that to induce the convergence of the iterative procedure. 
This can be done by employing optimized perturbation theory 
\cite{Yukalov_19,Yukalov_20,Yukalov_21,Yukalov_23}. For this purpose, it is 
possible to require that the sequence of approximations for observable quantities 
$\lgl\hat{A}\rgl_k$, related to the operator $\hat{A}$, be convergent. As a 
self-consistency condition, optimizing the choice, one can take  
\be
\label{32}
 \lgl \; \hat A \; \rgl_{k+1} - \lgl \; \hat A \; \rgl_k = 0 \; .
\ee 
For an operator of the form
\be
\label{33} 
 \hat A = \int \hat\psi^\dgr(\br) \hat A(\br) \hat\psi(\br) \; d\br \;  ,
\ee
the $k$-th order approximation is
\be
\label{34}
 \lgl \; \hat A \; \rgl_k = \pm i \lim_{\tau\ra +0} \int 
\int_{-\infty}^\infty e^{i\om\tau} \lim_{\br'\ra \br} \hat A(\br') 
G_k(\br,\br',\om)\; d\br \; \frac{d\om}{2\pi} \;  .
\ee
Then the optimization condition
\be
\label{35}
\lgl \; \hat A \; \rgl_1 - \lgl \; \hat A \; \rgl_0 = 0 
\ee
results in the equation
\be
\label{36}
\sum_{mn} \sum_{ij} \frac{n(\om_m)-n(\om_n)}{\om_m-\om_n} \; 
\Dlt_{ij}^{mn} A_{ji}^{nm} = 0 \;    .
\ee
The main contribution into sum (\ref{36}) is brought by the diagonal terms, which 
allows us to write
\be
\label{37}
 \sum_{nj} n(\om_n) [\; 1 \pm n(\om_n)\; ] \Dlt_{jj}^{nn} A_{jj}^{nn} = 0 \; .
\ee
In the single-band picture, considering only the lowest band, we get
\be
\label{38}
 \Dlt_{jj}^{00} = 0 \;  .
\ee
Explicitly, the latter equation writes as
$$
\int |\; w(\br-\br_j)\; |^2 \; U_0(\br) \; d\br =  
\int |\; w(\br-\br_j)\; |^2 \; U_L(\br) \; d\br \; +
$$
\be
\label{39}
 + \;
\int |\; w(\br-\br_j)\; |^2\; \Phi(\br-\br')\; \rho_0(\br')\; d\br d\br' \; ,
\ee
with the atomic density
\be
\label{40}
\rho_0(\br) = \sum_j \nu \; |\; w(\br-\br_j)\; |^2
\ee
and the filling factor
\be
\label{41}
 \nu \equiv \frac{N}{N_L} = \sum_n n(\om_n) \;  .
\ee
Taking into account the periodicity of the potentials $U_0$ and $U_L$ leads to 
the condition
\be
\label{42}
\int |\; w(\br) \; |^2 \; U_0(\br) \; d\br = 
\int |\; w(\br) \; |^2 \; U_L(\br) \; d\br \; + \;
\int |\; w(\br) \; |^2 \; \Phi(\br-\br') \; \rho_0(\br')\; d\br d\br' \;  .
\ee 

Thus when the optimization condition (\ref{38}), or (\ref{42}), is valid, the 
Green functions of zero and first order and, respectively, observable quantities 
of zero and first order are close to each other.

\section{Trial potential}

The optimization condition (\ref{42}) imposes a constraint on the choice of the 
trial potential $U_0$. For instance, in the vicinity of ${\bf r} \sim 0$, one can 
accept the potential of the form
\be
\label{43}
 U_0(\br) \simeq u_0 + \sum_{\al=1}^d \frac{m}{2} \; \ep_\al^2 r_\al^2 \;  ,
\ee
where $u_0$ is a static field and $\varepsilon_\alpha$ are effective frequencies 
of atomic oscillations. The related zero-order Wannier function can be approximated 
by the harmonic expression
\be
\label{44}
 w_n(\br) = \prod_\al \frac{(m\ep_\al/\pi)^{1/4}}{\sqrt{2^{n_\al}n_\al!}} \;
H_{n_\al}\left( \frac{r_\al}{l_\al} \right) 
\exp \left( -\;\frac{r_\al^2}{2l_\al^2} \right) \; ,
\ee
in which $n_\al=0,1,2,\ldots$ and 
$$
 l_\al \equiv \frac{1}{\sqrt{m\ep_\al} } \;  .
$$
The lowest-energy function (\ref{44}) is
\be
\label{45}
 w(\br) = \prod_\al \left( \frac{m\ep_\al}{\pi} \right)^{1/4}
\exp \left( -\;\frac{r_\al^2}{2l_\al^2} \right) \; .
\ee

In the vicinity of ${\bf r} \sim 0$, the optical-lattice potential can be written 
as
\be
\label{46}
U_L(\br) \simeq \sum_{\al=1}^d \frac{m}{2} \; \om_\al^2 r_\al^2 \qquad 
( r_\al \ra 0) \;   ,
\ee
with the frequency
\be
\label{47}
 \om_\al \equiv \frac{2\pi}{\lbd_\al} \; \sqrt{\frac{2}{m} \; U_\al} \;  .
\ee
Introducing the recoil energy
\be
\label{48}
 E_R^\al \equiv \frac{k_\al^2}{2m} \;  ,
\ee
for the optical-lattice frequency (\ref{47}), we have
\be
\label{49}
 \om_\al = 2\sqrt{E_R^\al U_\al} \;  .
\ee
Setting the static field
\be
\label{50}
 u_0 = \nu \sum_j \Phi(\br_j)   
\ee
yields the equation for the effective trial frequency $\ep_\al$,
$$
\frac{1}{4} \sum_\al \left( \ep_\al \; - \; \frac{\om_\al^2}{\ep_\al} \right)
+ \nu \sum_j \Phi(\br_j) \; =
$$
\be
\label{51}
 = \; \nu \sum_j 
\int |\; w(\br)\; |^2 \; \Phi(\br-\br')\; |\; w(\br'-\br_j)\;|^2\; d\br d\br' \; .
\ee

In general, we do not need to expand the interaction potential in powers of its 
coordinates. Then we would get the superharmonic approximation 
\cite{Yukalov_19,Yukalov_20,Yukalov_21} for the trial frequency $\ep_\al$. But, 
keeping in mind the properties of function (\ref{45}), it is admissible to expand 
the potential $\Phi(\br - \br')$ in powers of its variables. In that way, equation 
(\ref{51}) reduces to
\be
\label{52}
\frac{1}{4} \sum_\al \left( \ep_\al \; - \; \frac{\om_\al^2}{\ep_\al} \; - \;
\frac{\Om_\al^2}{\ep_\al} \right) = 0 \;  ,
\ee
in which
\be
\label{53}
 \Om_\al^2 \equiv \frac{2\nu}{m} \sum_j 
\frac{\prt^2\Phi(\br_j)}{\prt r_j^\al \prt r_j^\al} \;  .
\ee
Then we find the effective trial frequency
\be
\label{54}
 \ep_\al = \sqrt{\om_\al^2 + \Om_\al^2} \;  .
\ee
 
The meaning of the formula (\ref{54}) is rather clear. The effective oscillation 
frequency of atoms is defined by the optical-lattice potential as well as by atomic 
interactions. If the interactions are absent, the frequency is completely prescribed 
by the optical-lattice potential,
\be
\label{55}
\ep_\al = \om_\al \qquad ( \Om_\al = 0 ) \;   .
\ee
When the interactions are not zero, but yet rather weak, the effective frequency 
is close to that defined by the optical lattice, with small corrections caused by 
the interactions,
\be
\label{56}
 \ep_\al \simeq \om_\al \left( 1 + \frac{\Om_\al^2}{2\om_\al^2} \right) \qquad 
( \Om_\al \ll \om_\al ) \;  .
\ee
But when the interactions are strong, the effective oscillation frequency is 
mainly governed by atomic interactions, while the optical lattice provides small 
corrections,
\be
\label{57}
 \ep_\al \simeq \Om_\al \left( 1 + \frac{\om_\al^2}{2\Om_\al^2} \right) \qquad 
( \om_\al \ll \Om_\al ) \;  .
\ee
And if the interactions are so strong that the optical lattice becomes negligible, 
the oscillations are caused by atomic interactions,
\be
\label{58}
 \ep_\al = \Om_\al \qquad ( \om_\al = 0 ) \;   .
\ee
In the intermediate case, both the optical lattice as well as atomic interactions, 
regulate the effective frequency of atomic motion.

\section{Energy levels}

The energy levels for a many-body system can be defined in the following way. 
The equation of motion for Green functions can be written as
\be
\label{59}
 \int G^{-1}(\br,\br'',\om) G(\br'',\br',\om) \; d\br'' = \dlt(\br-\br') \;  ,
\ee
where the inverse propagator is
\be
\label{60}
 G^{-1}(\br,\br',\om) = (\om + \mu) \dlt(\br-\br') - H(\br,\br',\om) \; ,
\ee
with the effective Hamiltonian kernel
\be
\label{61}
 H(\br,\br',\om) = -\; \frac{\nabla^2}{2m} \; \dlt(\br-\br') + 
\Sigma(\br,\br',\om) \;  .
\ee

Introducing the effective Hamiltonian acting on Green functions by the formula
\be
\label{62}
\hat H(\br,\om)  G(\br,\br',\om)  = 
\int H(\br,\br'',\om) G(\br'',\br',\om)\; d\br'' \; ,
\ee
we come to the equation
\be
\label{63}
 [\; \om + \mu - \hat H(\br,\om)\; ]\; G(\br,\br',\om) = \dlt(\br-\br') \; .
\ee

When there exists a solution for the eigenproblem
\be
\label{64}
 \hat H(\br,\om) \vp_n(\br) = E_n\vp_n(\br) \;  ,
\ee
then and only then the Green function can be represented as the expansion
\be
\label{65}
 G(\br,\br',\om) = \sum_n G_n(\om) \vp_n(\br) \vp_n^*(\br') \;  ,
\ee
in which $G_n(\omega)$ is given by equation (\ref{28}). Equation (\ref{64}), 
called the effective Schr\"{o}dinger equation, defines the energy levels of 
a many-body system. The energy eigenvalues $E_n$ can be found by resorting to 
optimized perturbation theory \cite{Yukalov_23}. 

The zero approximation can be modeled by a diagonal Hamiltonian kernel
\be
\label{66}
 H_0(\br,\br',\om) = \hat H_0(\br) \dlt(\br - \br') \; ,
\ee
with the Hamiltonian
\be
\label{67}
 \hat H_0(\br)  = -\; \frac{\nabla^2}{2m} + U_0(\br)  
\ee
that is periodic over the lattice vectors
$$
 \hat H_0(\br + \br_j)  = \hat H_0(\br)  \; .
$$

However, Wannier functions are not eigenfunctions of this Hamiltonian (\ref{67}). 
The eigenfunctions of a periodic Hamiltonian are Bloch functions,
$$
\hat H_0(\br) \vp_{nk}(\br) = E_{nk} \vp_{nk}(\br) \;   .
$$
Expressing the Bloch functions through Wannier functions,
$$
 \vp_{nk}(\br) = \frac{1}{\sqrt{N_L}} 
\sum_j w_n(\br - \br_j) e^{i\bk\cdot\br_j} \;  ,
$$
we see that the action of the Hamiltonian on a Wannier function gives
\be
\label{68}
 \hat H_0(\br) w_n(\br - \br_j) = \sum_i E_{ij}^n w_n(\br - \br_i) \;  ,
\ee
where
$$
E_{ij}^n \equiv  \frac{1}{N_L} \sum_k E_{nk} e^{i\bk\cdot\br_{ij}} \;  .
$$
Equation (\ref{68}) can be transformed into
\be
\label{69}
 \hat H_0(\br) \psi_n(\br) = E_n \psi_n(\br) \;  ,
\ee
with
$$
\psi_n(\br) = \sum_j w_n(\br - \br_j) \; , \qquad 
E_n = \sum_j E_{ij}^n \;   .
$$
Hence the eigenfunctions of $\hat{H}_0$ are either Bloch functions or the sums 
$\psi_n({\bf r})$ of Wanier functions. 

But in the vicinity of ${\bf r} \approx {\bf r}_j$ equation (\ref{69}) takes the 
form
\be
\label{70}
 \hat H_0(\br) w_n(\br - \br_j) = E_n w_n(\br - \br_j) \; ,
\ee 
which means that Wannier functions are approximate solutions of this equation, in 
which $n = \{n_\alpha\}$ and $n_\alpha = 0,1,2,\ldots$. 

We accept the trial potential (\ref{43}), where the quantities $\ep_\al$ play the 
role of control functions. Then the zero-order energy levels are
\be
\label{71}
  E_n^{(0)} = u_0 + \sum_\al \left( n_\al + \frac{1}{2} \right) \ep_\al \; .
\ee
In the first order, equation (\ref{64}) yields the eigenvalues
\be
\label{72}
  E_n^{(1)} = \int w_n^*(\br) \hat H_1(\br) w_n(\br) \; d\br \; ,
\ee
where
\be
\label{73}
 \hat H_n^{(1)} = -\; \frac{\nabla^2}{2m} + U_1(\br)
\ee
and $U_1({\bf r})$ is defined by equation (\ref{26}). 

Considering only the ground-state energy level, we have the zero-order eigenvalue  
\be
\label{74}
 E_0^{(0)} = u_0 + \frac{1}{2} \sum_\al \ep_\al \;  .
\ee
Since
$$
 \int w(\br) \left( - \; \frac{\nabla^2}{2m} \right) w(\br)\; d\br = 
\frac{1}{4} \sum_\al \ep_\al \;  ,
$$
for the first-order approximation, we get
\be
\label{75}
 E_0^{(1)} = \frac{1}{4} \sum_\al \ep_\al + 
\int |\; w(\br)\; |^2 \; U_1(\br) \; d\br \;  .
\ee 
From here, we find
\be
\label{76}
 E_0^{(1)} = u_0 + \frac{1}{4} \sum_\al \left( \ep_\al + 
\frac{\om_\al^2}{\ep_\al} + \frac{\Om_\al^2}{\ep_\al} \right) \;  .
\ee

The trial frequency $\varepsilon_\alpha$ can be defined by the optimization condition
\be
\label{77}
 E_0^{(1)} - E_0^{(0)} = 0 \; ,
\ee
which results in the same expression (\ref{54}) for the frequency $\ep_\al$.

\section{Collective excitations}

In the previous sections, we have described the lowest energy state of an atomic 
system. This is actually the system ground state that is formed by an ensemble of 
atoms represented by harmonic oscillators, with a self-consistently defined frequency 
taking into account the presence of both, an optical lattice as well as of intersite 
atomic interactions. Collective excitations in such a system are known to be phonons 
\cite{Guyer_16,Brenig_24,Fredkin_25,Yukalov_26} that can be introduced in the 
following way.   
 
One treats the vectors ${\bf r}_j$ as the operators having the form
\be
\label{78}
 \br_j = \ba_j + \bu_j \;  ,
\ee
where ${\bf a}_j$ is a vector of an absolutely equilibrium effective lattice, while 
${\bf u}_j$ is a deviation operator, so that
\be
\label{79}
\ba_j \equiv \lgl \br_j \rgl \; , \qquad \lgl \bu_j \rgl = 0 \;   .
\ee
Also, it is convenient to introduce the relative quantities
$$
\br_{ij} \equiv \br_i - \br_j = \ba_{ij} - \bu_{ij} \; , \qquad
\ba_{ij} = \ba_i - \ba_j \; , \qquad  \bu_{ij} = \bu_i - \bu_j \;  .
$$

In the representation (\ref{10}) of the system Hamiltonian, one has
\be
\label{80}
 J_{ij} = J(\br_{ij}) \; , \qquad U_{ij} = U(\br_{ij}) \; , \qquad
U_{jj} = U(0) \equiv U \;  .
\ee
These quantities are expanded in powers of the deviations from the lattice sites 
${\bf a}_j$, keeping the terms up to the second order,
$$
J(\br_{ij}) \simeq J(\ba_{ij}) + \sum_\al J_{ij}^\al u_{ij}^\al \; - \; 
\frac{1}{2} \sum_{\al\bt} J_{ij}^{\al\bt} u_{ij}^\al u_{ij}^\bt \; , 
$$
\be
\label{81}
U(\br_{ij}) \simeq U(\ba_{ij}) + \sum_\al U_{ij}^\al u_{ij}^\al \; - \; 
\frac{1}{2} \sum_{\al\bt} U_{ij}^{\al\bt} u_{ij}^\al u_{ij}^\bt \;  ,
\ee
using the notations
$$
J_{ij}^\al \equiv \frac{\prt J(\ba_{ij})}{\prt a_i^\al} = 
\frac{\prt J(\ba_{ij})}{\prt a_{ij}^\al} \; , \qquad
U_{ij}^\al \equiv \frac{\prt U(\ba_{ij})}{\prt a_i^\al} = 
\frac{\prt U(\ba_{ij})}{\prt a_{ij}^\al} \; ,
$$
$$
J_{ij}^{\al\bt} \equiv \frac{\prt^2 J(\ba_{ij})}{\prt a_i^\al \prt a_j^\bt} = 
-\; \frac{\prt^2 J(\ba_{ij})}{\prt a_{ij}^\al \prt a_{ij}^\bt} \; , \qquad
U_{ij}^{\al\bt} \equiv \frac{\prt^2 U(\ba_{ij})}{\prt a_i^\al\prt a_j^\bt} = 
-\; \frac{\prt^2 U(\ba_{ij})}{\prt a_{ij}^\al \prt a_{ij}^\bt} \;  .
$$

Vibrational and atomic degrees of freedom can be decoupled according to the rule
$$
u_{ij}^\al u_{ij}^\bt \hat c_i^\dgr \hat c_j^\dgr \hat c_j \hat c_i = 
\lgl u_{ij}^\al u_{ij}^\bt \rgl \hat c_i^\dgr \hat c_j^\dgr \hat c_j \hat c_i + 
u_{ij}^\al u_{ij}^\bt \lgl \hat c_i^\dgr \hat c_j^\dgr \hat c_j \hat c_i \rgl - 
\lgl u_{ij}^\al u_{ij}^\bt \rgl \lgl \hat c_i^\dgr \hat c_j^\dgr \hat c_j \hat c_i \rgl \; ,
$$
$$
u_{ij}^\al u_{ij}^\bt \hat c_i^\dgr \hat c_j = 
\lgl u_{ij}^\al u_{ij}^\bt \rgl \hat c_i^\dgr \hat c_j +
u_{ij}^\al u_{ij}^\bt \lgl \hat c_i^\dgr \hat c_j \rgl -
\lgl u_{ij}^\al u_{ij}^\bt \rgl  \lgl \hat c_i^\dgr \hat c_j \rgl \; ,
$$
\be
\label{82}
\bp_j^2 \hat c_j^\dgr \hat c_j = \lgl \bp_j^2 \rgl  \hat c_j^\dgr \hat c_j +
\bp_j^2 \lgl \hat c_j^\dgr \hat c_j \rgl -
\lgl \bp_j^2 \rgl \lgl \hat c_j^\dgr \hat c_j \rgl \;  .
\ee

Then one can define the renormalized tunneling parameter
\be
\label{83} 
 \widetilde J_{ij} \equiv J(\ba_{ij} ) - \; \frac{1}{2} 
\sum_{\al\bt} J_{ij}^{\al\bt} \lgl u^\al_{ij} u^\bt_{ij} \rgl \; , 
\ee
renormalized atomic interactions
\be
\label{84}
 \widetilde U_{ij} \equiv U(\ba_{ij} ) - \; \frac{1}{2} 
\sum_{\al\bt} U_{ij}^{\al\bt} \lgl u^\al_{ij} u^\bt_{ij} \rgl \;  ,
\ee
renormalized dynamic interactions
\be
\label{85}
 \Phi_{ij}^{\al\bt} \equiv U_{ij}^{\al\bt} \;
\lgl \hat c_i^\dgr \hat c_j^\dgr \hat c_j \hat c_i \rgl -
 2J_{ij}^{\al\bt} \; \lgl \hat c_i^\dgr \hat c_j \rgl \; ,
\ee
and the effective force
\be
\label{86}
 F_{ij}^\al = 2J_{ij}^\al \; \hat c_i^\dgr \hat c_j - 
U_{ij}^\al \; \hat c_i^\dgr \hat c_j^\dgr  \hat c_j \hat c_i \;  .
\ee
The filling factor (\ref{41}) becomes
\be
\label{87}
 \nu \equiv \frac{N}{N_L} = 
\frac{1}{N_L} \sum_j \; \lgl \hat c_j^\dgr \hat c_j \rgl  =
 \lgl \hat c_j^\dgr \hat c_j \rgl \; .
\ee

In that way, we come to the Hamiltonian
\be
\label{88} 
 \hat H = E_N + \hat H_{at} + \hat H_{vib} + \hat H_{def} \;  .
\ee
The first is the nonoperator term
\be
\label{89}
 E_N = \frac{1}{4} \sum_{i\neq j} 
\sum_{\al\bt} \Phi_{ij}^{\al\bt} \lgl u_{ij}^\al u_{ij}^\bt \rgl - 
\nu \sum_j \frac{\lgl \bp_j^2\rgl}{2m} \;  .
\ee
The second is the renormalized atomic term
$$
\hat H_{at} = - \sum_{i\neq j} \widetilde J_{ij} \hat c_i^\dgr \hat c_j \; + \;
\frac{U}{2} \sum_j \hat c_j^\dgr \hat c_j^\dgr \hat c_j \hat c_j \; +
$$
\be
\label{90}
  + \; \frac{1}{2} \sum_{i\neq j} 
\widetilde U_{ij} \hat c_i^\dgr \hat c_j^\dgr \hat c_j \hat c_i \; + \;
\sum_j \left( \frac{\lgl\bp_j^2\rgl}{2m} + U_L \right) \hat c_j^\dgr \hat c_j \; .
\ee
The third is the term with vibrational degrees of freedom
\be
\label{91}
 \hat H_{vib} = \nu \sum_j \frac{\bp_j^2}{2m} \; - \; 
\frac{1}{4} \sum_{i\neq j} 
\sum_{\al\bt} \Phi_{ij}^{\al\bt} u_{ij}^\al u_{ij}^\bt \;  .
\ee
Substituting here ${\bf a}_{ij} = {\bf a}_i - {\bf a}_j$ yields 
\be
\label{92}
 \hat H_{vib} = \nu \sum_j \frac{\bp_j^2}{2m} \; + \; 
\frac{1}{4} \sum_{i\neq j} 
\sum_{\al\bt} \Phi_{ij}^{\al\bt} u_i^\al u_j^\bt \;  .
\ee
And the last term 
\be
\label{93}
\hat H_{def} = -\; \frac{1}{2} \sum_{i\neq j} \sum_\al F_{ij}^\al u_{ij}^\al
\ee
describes local deformations of the lattice. Introducing the notation for the 
effective force acting on a $j$-th atom
\be
\label{94}
 F_j^\al \equiv \sum_{i(\neq j)} F_{ji}^\al \;  ,
\ee
with the properties
\be
\label{95}
F_{ji}^\al = - ( F_{ij}^\al )^+ \; , \qquad \lgl F_i^\al \rgl = 0 \;  ,
\ee
we obtain
\be
\label{96}
\hat H_{def} = -\; \frac{1}{2} \sum_j 
\sum_\al \left [ F_j^\al + ( F_j^\al )^+ \right] u_j^\al \;  .
\ee   

Following the method of decoupling of atomic and vibrational degrees of freedom, 
we have
\be
\label{97}
 F_j^\al u_j^\al = F_j^\al \lgl u_j^\al \rgl + \lgl F_j^\al \rgl u_j^\al
-  \lgl F_j^\al \rgl \lgl u_j^\al \rgl \;  .
\ee
But, since $\lgl F_j^\al \rgl=0$ and $\lgl u_j^\al \rgl=0$, we come to the 
conclusion that the deformation term is negligible, $\hat{H}_{def} = 0$. 

At the next step, we introduce phonon operators $b_{ks}$ by the canonical 
transformation
$$
\bu_j = \vec{\dlt}_j + \frac{1}{\sqrt{2N}} 
\sum_{ks} \sqrt{ \frac{\nu}{m\om_{ks}} } \;
\bfe_{ks} \left( b_{ks} + b^\dgr_{-ks} \right) e^{i\bk\cdot\ba_j} \; ,
$$
\be
\label{98}
 \bp_j = -\; \frac{i}{\sqrt{2N}} 
\sum_{ks} \sqrt{ \frac{m\om_{ks}}{\nu} } \;
\bfe_{ks} \left( b_{ks} - b^\dgr_{-ks} \right) e^{i\bk\cdot\ba_j} \;  ,
\ee
in which $k$ is momentum, $s$ is a polarization label, ${\bf e}_{ks}$ are the 
polarization vectors with the properties
$$
 \bfe_{ks} = \bfe_{-ks} \; , \qquad 
\sum_s e_{ks}^\al e_{ks}^\bt = \dlt_{\al\bt} \; , \qquad
\bfe_{ks} \cdot \bfe_{ks'} = \sum_\al e_{ks}^\al e_{ks'}^\al = \dlt_{ss'} \; ,
$$ 
and the phonon spectrum $\omega_{ks} = \omega_{-ks}$ is defined by the eigenproblem
\be
\label{99}
\frac{\nu}{m} \sum_{j(\neq i)} 
\sum_\bt \Phi_{ij}^{\al\bt} e^{i\bk\cdot\ba_{ij}} e_{ks}^\bt =
\om_{ks}^2 e_{ks}^\al \;  .
\ee
The first term in the first of equations (\ref{98}) is responsible for the 
diagonalization of the phonon Hamiltonian, if the deformation part is taken into 
account, which requires the equality
$$
 \dlt_j^\al = \sum_{i(\neq j)} \sum_\bt \gm_{ji}^{\al\bt} F_i^\bt \;  ,
$$
with
$$
 \gm_{ij}^{\al\bt} \equiv \frac{\nu}{m} 
\sum_{ks} \frac{e_{ks}^\al e_{ks}^\bt}{m\om_{ks}^2} \; e^{i\bk\cdot\ba_{ij}} \; .
$$
On the average,  
$$
\lgl \dlt_j^\al \rgl = 0 \; .
$$

Thus Hamiltonian (\ref{88}) is transformed into
\be
\label{100}
 \hat H = E_N + \hat H_{at} + \hat H_{ph} + \hat H_{ind} \;  ,
\ee
where $E_N$ and $\hat{H}_{at}$ are given by equations (\ref{89}) and (\ref{90}). 
The phonon Hamiltonian is
\be
\label{101}
 \hat H_{ph} = \sum_{ks} \om_{ks} 
\left( b_{ks}^\dgr b_{ks} + \frac{1}{2} \right) \;  ,
\ee
and the term
\be
\label{102}
\hat H_{ind} = \sum_{i\neq j} \sum_{\al\bt} F_i^\al \gm_{ij}^{\al\bt} F_j^\bt
\ee
characterizes additional atomic interactions induced by local deformations caused 
by phonon exchange. This term does not arise, if the deformation Hamiltonian (\ref{96}) 
has been neglected, as is explained above. The induced term (\ref{102}) also is zero, 
if we involve the decoupling
$$
F_i^\al F_j^\bt = F_i^\al \lgl F_j^\bt \rgl + \lgl F_i^\al \rgl F_j^\bt -
 \lgl F_i^\al \rgl \lgl F_j^\bt \rgl \; ,
$$
since $\lgl F_j^\al\rgl=0$.

\section{Average quantities}

In this way, the atomic system, including collective phonon excitations, is described 
by the Hamiltonian
\be
\label{103}
 \hat H = E_N + \hat H_{at} + \hat H_{ph} \;  ,
\ee
in which the atomic and phonon degrees of freedom are separated. Respectively, the 
Hilbert space of the system is represented by the tensor product
\be
\label{104}
 \cH = \cH_{at} \bigotimes \cH_{ph}  
\ee
of the Hilbert spaces corresponding to atomic and phonon degrees of freedom. Then 
the partition function takes the form
\be
\label{105}
 {\rm Tr}_\cH e^{-\bt\hat H} = 
e^{-\bt E_N} {\rm Tr}_{\cH_{at}} e^{-\bt\hat H_{at}} 
{\rm Tr}_{\cH_{ph}} e^{-\bt\hat H_{ph}} \;  .
\ee

The averages of the operators, acting on either the atomic Hilbert space or on the 
phonon Hilbert space, are calculated according to the rule
$$
\lgl \hat A_{at} \rgl = 
\frac{{\rm Tr}_\cH\hat A_{at}e^{-\bt\hat H}}{{\rm Tr}_\cH e^{-\bt\hat H}} =
\frac{{\rm Tr}_{\cH_{at}}\hat A_{at}e^{-\bt\hat H_{at}}}
{{\rm Tr}_{\cH_{at}} e^{-\bt\hat H_{at}}}  \; ,
$$
\be
\label{106}
 \lgl \hat A_{ph} \rgl = 
\frac{{\rm Tr}_\cH\hat A_{ph}e^{-\bt\hat H}}{{\rm Tr}_\cH e^{-\bt\hat H}} =
\frac{{\rm Tr}_{\cH_{ph}}\hat A_{ph}e^{-\bt\hat H_{ph}}}
{{\rm Tr}_{\cH_{ph}} e^{-\bt\hat H_{ph}}}  \; .
\ee
Thus we find the averages
$$
\lgl u_{ij}^\al u_{ij}^\bt \rgl = 
2 (1 - \dlt_{ij} ) \lgl u_j^\al u_j^\bt \rgl \; ,
\qquad
\lgl u_i^\al u_j^\bt \rgl = \dlt_{ij} \; \frac{\nu}{2N} \sum_{ks} 
\frac{e_{ks}^\al e_{ks}^\bt}{m\om_{ks}} \; 
\coth \left( \frac{\om_{ks}}{2T} \right) \; ,
$$
\be
\label{107}
 \frac{\lgl \bp_j^2 \rgl}{2m} = \frac{1}{4\nu N} 
\sum_{ks} \om_{ks}  \coth \left( \frac{\om_{ks}}{2T} \right) \;  ,
\ee
where the phonon frequency is given by the expression
\be
\label{108}
 \om_{ks}^2 = \frac{\nu}{m} \sum_{j(\neq i)} 
\sum_{\al\bt} \Phi_{ij}^{\al\bt} e_{ks}^\al e_{ka}^\bt e^{i\bk\cdot\ba_{ij}} \; .
\ee

An important quantity characterizing the atomic localization is the mean square 
deviation
\be
\label{109}
 r_0^2 \equiv \sum_{\al=1}^d  \lgl u_j^\al u_j^\al \rgl \;  .
\ee
Assuming an ideal lattice implies that
\be
\label{110}
 \lgl u_j^\al u_j^\bt \rgl = \dlt_{\al\bt} \; \frac{r_0^2}{d} \;  .
\ee
   
In conclusion to this section, let us show how the parameters, entering the 
above formulas, are calculated. Consider the effective average interaction 
between different lattice sites
$$
U(\ba_{ij} ) = \int |\; w(\br-\ba_i )\; |^2 \;\Phi(\br-\br') \; |\; 
w(\br'-\ba_j )\; |^2 \; d\br d\br' \; =
$$
\be
\label{111}
  = \;
\int w^2(\br) w^2(\br') \Phi(\br-\br'+ \ba_{ij} ) \; d\br d\br'
\ee
and the matrix
\be
\label{112}
 U_{ij}^{\al\bt}  =  \int w^2(\br) w^2(\br') \; 
\frac{\prt^2}{\prt a_i^\al \prt a_j^\bt} \;  
\Phi(\br-\br'+ \ba_{ij} ) \; d\br d\br' .
\ee
Using the expansion
$$
\Phi(\br-\br'+ \ba ) \simeq \Phi(\ba) + 
\sum_\al \frac{\prt \Phi(\ba)}{\prt a_\al} \; ( r_\al - r_\al' ) + 
\frac{1}{2} \sum_{\al\bt} \; \frac{\prt^2 \Phi(\ba)}{\prt a_\al \prt a_\bt} \; 
(r_\al r_\bt - 2 r_\al r_\bt' + r_\al' r_\bt' ) \;  ,
$$   
we find
\be
\label{113}
U(\ba_{ij}) \simeq \Phi(\ba_{ij}) + 
\sum_\al \frac{\prt^2 \Phi(\ba_{ij})}{\prt a_i^\al \prt a_i^\al} \; l_\al^2 \; ,
\ee
and 
\be
\label{114}
U_{ij}^{\al\bt} \equiv \frac{\prt^2 U(\ba_{ij})}{\prt a_i^\al \prt a_j^\bt}
\approx \frac{\prt^2 \Phi(\ba_{ij})}{\prt a_i^\al \prt a_j^\bt} \; .
\ee
Using the similar expansions, we also have
$$
J(\ba_{ij}) = \sum_\al \left\{ \frac{\ep_\al}{8} 
\left[ \left( \frac{a_{ij}^\al}{l_\al}\right)^2 - 2 \right] - U_\al \right\}\;
 \exp \left\{ - \; \frac{1}{4} 
\sum_\al \left( \frac{a_{ij}^\al}{l_\al}\right)^2 \right\} \; .
$$

\section{Mean square deviation}

The mean square deviation (\ref{109}) is an important quantity characterizing 
the stability of localized states. To be more concrete, we shall use the Debye 
approximation that is known to be a good description of quantum lattice structures 
\cite{Guyer_16} and optical lattices \cite{Yukalov_26}. 

Defining the effective average frequency by the equation 
\be
\label{115}
 \om_k^2 = \frac{1}{d} \sum_{s=1}^d \om_{ks}^2 \;  ,
\ee
we get
\be
\label{116}
 \om_k^2 = -\; \frac{\nu}{m} \sum_{j(\neq i)} D_{ij} e^{i\bk\cdot\ba_{ij}} \; ,
\ee
with the dynamical matrix
\be
\label{117}
 D_{ij} \equiv -\; \frac{1}{d} \sum_{\al=1}^d \Phi_{ij}^{\al\al} \qquad
( i \neq j ) \; .
\ee
Keeping in mind that atoms are well localized in their lattice sites means that 
$\hat{c}^\dagger_i \hat{c}_j = \nu \delta_{ij}$. Then we have
\be
\label{118}
 \Phi_{ij}^{\al\bt}  = \nu^2 U_{ij}^{\al\bt} = 
\nu^2 \; \frac{\prt^2 U(\ba_{ij})}{\prt a_i^\al\prt a_j^\bt} \; ,
\ee
which gives the dynamical matrix
\be
\label{119}
 D_{ij} = - \; \frac{\nu^2}{d} 
\sum_{\al=1}^d 
\frac{\prt^2 U(\ba_{ij})}{\prt a_i^\al\prt a_j^\al} =
\frac{\nu^2}{d} \sum_{\al=1}^d 
\frac{\prt^2 U(\ba_{ij})}{\prt a_{ij}^\al\prt a_{ij}^\al} \;  .
\ee
In this approximation, the mean square deviation (\ref{109}) becomes
\be
\label{120}
r_0^2 = \frac{\nu d}{2m\rho} \int_\mathbb{B} \frac{1}{\om_k} \;
\coth \left( \frac{\om_k}{2T} \right) \frac{d\bk}{(2\pi)^d} \;  ,
\ee
with the integration over the Brillouin zone.

Taking, for concreteness, a cubic lattice with the nearest-neighbour distance $a$, 
we have for the frequency (\ref{116}) the expression
\be
\label{121}
 \om_k^2 = \frac{4\nu}{m} \; D 
\sum_{\al=1}^d \sin^2 \left( \frac{k_\al a}{2} \right) \;  ,
\ee
where the dynamic parameter is
\be
\label{122}
 D  = \frac{\nu^2}{d} \sum_{\al=1}^d 
\frac{\prt^2 U(\ba)}{\prt a^\al\prt a^\al} \;  .
\ee
In the long-wave limit, the spectrum is of the phonon type,
\be
\label{123}
\om_k \simeq c_0 k \qquad 
\left( k^2 = \sum_{\al=1}^d k_\al^2 ~ \ra ~ 0 \right) \;  ,
\ee
with the sound velocity
\be
\label{124}
c_0 = \sqrt{ \frac{\nu}{m} \; D a^2 } \; .
\ee

In the Debye approximation, one replaces the integration over the Brillouin zone by 
the integration over the Debye sphere, so that
\be
\label{125}
\int_\mathbb{B} \frac{d\bk}{(2\pi)^d} ~ \ra ~
\frac{2}{(4\pi)^{d/2}\Gm(d/2)} \int_0^{k_D} k^{d-1}\; dk \;   .
\ee
The Debye radius $k_D$ is found from the normalization
\be
\label{126}
\int_\mathbb{B} \frac{d\bk}{(2\pi)^d}  = \frac{N_L}{V} = \frac{\rho}{\nu} \; ,
\ee
in which $\rho$ is average density
\be
\label{127}
 \rho \equiv \frac{N}{V} = \frac{\nu}{a^d} \;  .
\ee
This gives the Debye radius
\be
\label{128}
k_D = \frac{\sqrt{4\pi}}{a} \; 
\left[ \frac{d}{2} \; \Gm\left( \frac{d}{2} \right) \right]^{1/d} \; .
\ee
For particular dimensionalities, we have
$$
k_D = \frac{\pi}{a} \qquad ( d = 1 ) \; ,
$$
$$
k_D = \frac{2\sqrt{\pi}}{a} \qquad ( d = 2 ) \; ,
$$
\be
\label{129}
k_D = \frac{(6\pi^2)^{1/3}}{a} \qquad ( d = 3 ) \; .
\ee

One accepts the phonon spectrum (\ref{123}),
\be
\label{130}
\om_k = c_0 k \qquad ( 0 \leq k \leq k_D )
\ee
that is cut from above by the Debye temperature
\be
\label{131}
T_D  = c_0 k_D = \sqrt{4\pi \; \frac{\nu D}{ m} } \; 
\left[ \frac{d}{2}\; \Gm\left( \frac{d}{2} \right) \right]^{1/d} \; .
\ee
Thus for different dimensionalities, one has
$$
T_D = \pi\; \sqrt{\frac{\nu}{m}\; D} \qquad ( d = 1 ) \; ,
$$
$$
T_D =  \sqrt{4\pi\;\frac{\nu}{m}\; D} \qquad ( d = 2 ) \; ,
$$
\be
\label{132}
 T_D = (6\pi^2 )^{1/3} \; \sqrt{\frac{\nu}{m}\; D} \qquad ( d = 3 ) \;  .
\ee

Replacing the integration according to the rule
$$
\int_\mathbb{B} \frac{d\bk}{(2\pi)^d} ~ \ra ~
\frac{d}{(k_D a)^d} \int_0^{k_D} k^{d-1}\; dk 
$$
results in the mean square deviation
\be
\label{133}
 r_0^2 = \frac{d}{2mT_D} 
\int_0^1 x^{d-2} \coth\left( \frac{T_D}{2T}\; x\right) \; dx \;  .
\ee
At zero temperature, this yields
\be
\label{134}
  r_0^2 = \frac{d^2}{2(d-1)mT_D} \qquad ( T = 0 ) \; ,
\ee 
while at high temperature, we get
\be
\label{135}
  r_0^2 \simeq \frac{Td^2}{(d-2)mT_D^2} \qquad ( T \gg T_D ) \;  .
\ee

\section{Stability of localized states}

The stability of a localized state can be characterized by the Lindemann criterion 
\cite{Lindemann_27}, according to which the mean square deviation has to be smaller
than half of the distance between the nearest neighbours,
\be
\label{136}
 \frac{r_0}{a} < \frac{1}{2} \; ,
\ee
otherwise strong atomic oscillations destroy the localization. With the mean square 
deviation (\ref{133}), this criterion reads as
\be
\label{137}
\frac{12E_R}{\pi^2 T_D d} 
\int_0^1 x^{d-2} \coth\left( \frac{T_D}{2T}\; x\right) \; dx < 1 \;  ,
\ee
where the recoil energy
\be
\label{138}
E_R \equiv \frac{k_0^2}{2m} = \frac{\pi^2 d}{2m a^2}
\ee
is introduced. 

At zero temperature, the stability condition takes the form
\be
\label{139}
 \frac{T_D}{E_R} > \frac{4d}{\pi^2(d-1)} \qquad ( T = 0 ) \; .
\ee
Hence, one-dimensional periodic localized structures are not stable at zero 
temperature. While for higher dimensionalities, the stability condition gives
$$
 \frac{T_D}{E_R} > \frac{8}{\pi^2} \qquad ( d = 2 \; , ~ T = 0 ) \; .
$$
\be
\label{140}
 \frac{T_D}{E_R} > \frac{6}{\pi^2} \qquad ( d = 3 \; , ~ T = 0 ) \; .
\ee
For temperatures above $T_D$, the stability condition is
\be
\label{141}
 \frac{T_D}{E_R} > \frac{8d}{\pi^2(d-2)} \qquad ( T \gg T_D ) \;  .
\ee
At these temperatures, two-dimensional localized lattices are not stable. And the 
stability condition for a three-dimensional localized lattice is
\be
\label{142}
 \frac{T_D}{E_R} > \frac{24}{\pi^2} \qquad ( d = 3 \; , ~ T \gg T_D ) \;  .
\ee

It is possible to notice that the Debye temperature is very close to the frequency 
$\Omega_\alpha$ in definition (\ref{53}). Really, their ratio is
\be
\label{143}
 \frac{T_D}{\Om_\al}  = \nu \; \sqrt{\frac{\pi}{d} } \; \left[ \frac{d}{2} \;
\Gm\left( \frac{d}{2} \right) \right]^{1/d}  \; .
\ee
Taking for an estimate $d=3$, the number of the nearest neighbors $2d=6$, and 
the filling factor $\nu = 1$, we get $T_D = 1.1 \Omega_\alpha$.  

In this way, the localized state is stable, when the interactions between 
different lattice sites are sufficiently strong, such that the Debye temperature 
$T_D$ is essentially larger than the recoil energy $E_R$, with the above localization 
conditions be valid. When the latter do not hold, the system is not localized.

\section{Classification of states}

The character of a state realized in the system depends on the space dimensionality 
and the interplay between three types of the parameters, the optical-lattice 
frequencies that can be represented as
\be
\label{144}
 \om_\al = \sqrt{\frac{4}{3}\; E_R U_\al} \;  ,
\ee
the recoil energy $E_R$, and the Debye temperature $T_D$. The effective frequency 
of atomic oscillations (\ref{54}) is regulated by both the optical-lattice frequency 
$\omega_\alpha$ and the Debye temperature $T_D$. As is clear from expressions 
(\ref{55}) to (\ref{58}), the atomic frequency $\ep_\al$ varies from $\om_\al$ at 
weak intersite interactions to the Debye temperature $T_D$ under strong interactions 
between different lattice sites.  

The behaviour of optical lattices, where there are no intersite interactions, 
hence $T_D=0$, is well known, as can be inferred from the reviews 
\cite{Morsch_1,Moseley_2,Yukalov_3} and books \cite{Pethick_4,Ueda_5}. At zero 
temperature, integer filling factor, and deep lattice, where $U_\al\gg E_R$, one 
has the Mott insulator, while in the opposite case of a shallow lattice, when 
$U_\al\ll E_R$, the optical lattice is conducting. If either temperatures are
finite or the filling factor is not integer, the lattice cannot be completely 
localized, but is conducting. The inclusion of the intersite interactions makes 
the overall picture much more diversified. Summarizing the available knowledge 
on optical lattices without intersite interactions 
\cite{Morsch_1,Moseley_2,Yukalov_3,Pethick_4,Ueda_5} and the results obtained 
in the previous sections, we have the following picture, where we keep in mind 
a three-dimensional system $(d=3)$. 

Consider first the case of a deep optical lattice, such that
\be
\label{145}
 \frac{\om_\al}{E_R}  \gg 1 \qquad 
\left( \frac{U_\al}{E_R} \gg 1 \right) \; .
\ee
If there are intersite interactions, so that $T_D > 0$, but they are rather weak, 
\be
\label{146}
 0 < T_D \ll E_R \qquad (conducting \; optical \; lattice) \;  ,
\ee
we have a conducting optical lattice, with the lattice vector created by the 
applied laser beams. For intersite interactions much stronger than $E_R$, but 
smaller than $\omega_\alpha$, the optical lattice becomes insulating,
\be
\label{147} 
E_R \ll T_D \ll \om_\al \qquad (insulating \; optical \; lattice) \;  .
\ee
For even stronger intersite interactions, the insulating optical lattice 
transforms into a localized quantum crystal, with the lattice vector prescribed 
by the minimum of free energy,
\be
\label{148}
E_R \ll \om_\al \ll T_D \qquad (localized \; quantum \; crystal) \;  .
\ee

If the imposed optical lattice is shallow, with $U_\alpha \ll E_R$, when
\be
\label{149}
 \frac{\om_\al}{E_R} \ll 1 \qquad 
\left( \frac{U_\al}{E_R} \ll 1 \right) \;  ,
\ee
then, depending on the strength of the intersite interactions, there can occur 
the following states. Under quite weak intersite interactions, when
\be
\label{150}
 0 \leq T_D \ll \om_\al \qquad (conducting \; optical \; lattice) \;  ,
\ee
we have a conducting optical lattice. When the intersite interactions become 
prevailing over the optical-lattice potential, the system properties are governed 
by these interactions, but they may yet be not sufficient for localizing atoms, so 
that when
\be
\label{151}
 \om_\al \ll T_D \ll E_R \qquad (delocalized \; quantum \; crystal) \;  ,
\ee
there happens a delocalized quantum crystal. And when the intersite interactions 
are lager than all other potentials, 
\be
\label{152}
 \om_\al \ll E_R \ll T_D \qquad (localized \; quantum \; crystal) \;  ,
\ee
the quantum crystal localizes. Recall that this classification concerns 
three-dimensional systems.

\section{Low-dimensional finite localized lattices}

As is shown in Sec. 9, one-dimensional localized lattices cannot exist at any 
temperature, including zero temperature, and two-dimensional localized lattices 
cannot exist at finite temperatures. These conclusions have been obtained for 
macroscopic periodic systems, for which the minimal wave-vector in the Brillouin 
zone is zero. But if the system is finite, the minimal wave-vector is limited 
because of the finite-size effects, being of the order of 
\be
\label{153}
 k_{min} = \frac{\pi}{L} \qquad \left( L = N_L^{1/d} a \right) \;  .
\ee
Then the integral in equation (\ref{133}) is limited from below by the quantity
\be
\label{154}
 x_{min} \equiv \frac{k_{min}}{k_D} = \frac{\pi}{k_D a N_L^{1/d}}  
\ee
that is asymptotically small with respect to $N_L$, but finite,
$$
x_{min} = \frac{1}{N_L} \qquad ( d = 1) \; ,
$$
$$
x_{min} = \frac{1}{2} \; \sqrt{\frac{\pi}{N_L} } \qquad ( d = 2) \; .
$$

For a one-dimensional system at zero temperature, equation (\ref{133}) gives
\be
\label{155}
 r_0^2 \cong \frac{\ln N_L}{2mT_D}  \qquad ( d = 1 \; , ~ T = 0) \; ,
\ee
while for a two-dimensional system $r_0^2 = 2/mT_D$. This means that 
two-dimensional localized lattices can be of any size, while the size of 
one-dimensional localized lattices is restricted by the inequality
\be
\label{156}
  N_L < \exp\left( \frac{\pi^2T_D}{4E_R} \right) \qquad 
( d = 1 \; , ~ T = 0) \; ,
\ee
following from the Lindemann stability condition (\ref{136}).

At finite temperature, equation (\ref{133}) results in the expressions
$$
 r_0^2 \cong \frac{T N_L}{mT_D^2}  \qquad ( d = 1 \; , ~ T > 0) \;
$$
\be
\label{157}
  r_0^2 \cong \frac{T\ln N_L}{mT_D^2}  \qquad ( d = 2 \; , ~ T > 0) \;  .
\ee
Therefore, the Lindemann criterion imposes the constraints
$$
 N_L < \frac{\pi^2 T_D^2}{8TE_R}  \qquad ( d = 1 \; , ~ T > 0) \; ,
$$
\be
\label{158}   
N_L < \exp\left( \frac{\pi^2 T_D^2}{4TE_R} \right) \qquad 
( d = 2 \; , ~ T > 0) \;
\ee
on the size of localized lattices. Such a size, depending on the parameters, 
can be quite large. Take, for instance, $T \sim T_D \sim 10 E_R$. Then finite 
localized lattices can contain the number of lattice sites as $N_L<10$ for 
$d=1$ and $N_L<5\times 10^{10}$ for $d=2$. 

Thus one- and two-dimensional localized structures can exist as finite objects, 
with the sizes prescribed by temperature, recoil energy, and the strength of 
intersite interactions characterized by the Debye temperature. Such finite 
localized structures can exist even if the corresponding macroscopic objects 
are not localized.

\section{Conclusion}

In this paper, we consider optical lattices of atoms or molecules that, 
in addition to one-site interactions, exhibit interactions between different 
lattice sites. These intersite interactions can be rather strong, inducing 
collective phonon excitations above the ground-state energy level. The 
properties of optical lattices in the case of absent intersite interactions 
are known. We pose the question: What happens if the strength of the intersite 
interactions increases, from the situation where this strength is much weaker 
than the optical lattice potential and the system properties are governed by 
the optical lattice to the case where the strength of the intersite interactions 
is much larger than the optical potential barrier? 

The properties of the system are shown to be mainly regulated by the three types 
of the system parameters, optical-lattice frequencies, recoil energy, and the 
Debye temperature characterizing the strength of the intersite atomic interactions. 
Depending on the interplay between these parameters, in three dimensions, there can 
exist conducting optical lattices, insulating optical lattices, delocalized quantum 
crystals, and localized quantum crystals.

In one-dimensional systems, localized macroscopic lattice structures cannot arise 
at any temperature, either zero or not. And in two dimensions, localized lattices 
can arise only at zero temperature. Nevertheless, one-dimensional localized chains 
and two-dimensional plains of localized atoms can exist at zero as well as at 
nonzero temperature, provided such low-dimensional structures are of finite sizes. 

The consideration in the paper concerns equilibrium systems. It is also possible 
to mention that even when an equilibrium system cannot occur, there can be created 
a similar metastable system with sufficiently long lifetime. Thus, assume that an 
insulating lattice is unstable as an equilibrium system. Anyway, it can be created 
as a metastable object with the lifetime
$$
t_{met} = \min_\al \; 
\frac{2\pi}{\om_\al} \; \exp\left( \frac{U_\al}{E_R} \right) \; .
$$

The existence of long-lived metastable systems, as well as the occurrence of 
stable finite low-dimensional localized lattices of chains and planes, explains 
why such structures can be created and studied in experiments. In the present 
paper, we have considered the destiny of optical lattices, loaded by atoms or 
molecules with strong intersite interactions inducing collective excitations. 
Similar investigations could be appropriate for other finite quantum systems, 
whose stability, and hence existence, are intimately connected with the presence 
of collective excitations \cite{Birman_28}.

\end{document}